%% file: main.tex
\documentclass[conference]{IEEEtran}
\IEEEoverridecommandlockouts
\usepackage{url} 
\usepackage{cite}
\usepackage{amsmath,amssymb,amsfonts}
\usepackage{graphicx}
\usepackage{textcomp}
\usepackage{xcolor}
\usepackage{algorithm}
\usepackage{algpseudocode}
\usepackage{booktabs}
\usepackage{multirow}
\usepackage{rotating}
\usepackage{tikz}
\usepackage{array}
\usepackage{tabularx}
\usepackage{enumitem}
\usepackage{comment}
\usepackage{xspace}
\usetikzlibrary{shapes.geometric, arrows.meta, positioning, fit, backgrounds, decorations.pathreplacing}

\newcommand{\FACET}{\emph{FAME}\xspace}

\definecolor{cDeepBlue}{HTML}{1A5276}
\definecolor{cAccBlue}{HTML}{2E86C1}
\definecolor{cLtBlue}{HTML}{D6EAF8}
\definecolor{cTeal}{HTML}{148F77}
\definecolor{cLtTeal}{HTML}{D1F2EB}
\definecolor{cDeepRed}{HTML}{922B21}
\definecolor{cLtRed}{HTML}{FADBD8}
\definecolor{cAmber}{HTML}{D68910}
\definecolor{cLtAmber}{HTML}{FDEBD0}
\definecolor{cCharcoal}{HTML}{2C3E50}
\definecolor{cGray}{HTML}{7F8C8D}
\definecolor{cLtGray}{HTML}{F2F3F4}
\definecolor{cBorder}{HTML}{BDC3C7}

\def\BibTeX{{\rm B\kern-.05em{\sc i\kern-.025em b}\kern-.08em
    T\kern-.1667em\lower.7ex\hbox{E}\kern-.125emX}}

\begin{document}

\title{FAME: Failure-Aware Mixture-of-Experts for Message-Level Log Anomaly Detection}

\author{
\IEEEauthorblockN{Huanchi Wang, Zihang Huang, Yifang Tian, Kristina Dzeparoska, Hans-Arno Jacobsen and Alberto Leon-Garcia}
\IEEEauthorblockA{
  \textit{Department of Electrical and Computer Engineering}\\
  University of Toronto, Toronto, Ontario\\
  \{huanchi.wang, zihang.huang, yifang.tian,kristina.dzeparoska\}@mail.utoronto.ca, \\jacobsen@eecg.toronto.edu, 
  alberto.leongarcia@utoronto.ca}

}

\maketitle
\pagestyle{plain}

\input{abstract_final}

\input{introduction_final}

\input{Preliminary_final}


\input{methodology_final}

\input{evaluation_final}

\input{case_study_final}
\input{background_final}
\input{limitations_final}
\input{conclusion_final}

\bibliographystyle{IEEEtran}
\bibliography{references}

\end{document}

%% file: abstract_final.tex
\begin{abstract}
Production systems generate millions of log lines daily, and many operational anomaly detectors still operate at the window level, flagging groups of lines rather than identifying the specific message responsible. This coarse granularity forces operators to inspect many routine lines per alert. Message-level detection offers finer granularity, but remains challenging. A single event template may correspond to both normal and anomalous messages, failures arise from heterogeneous subsystems, and line-level labeling at scale is impractical. Although large language models (LLMs) can reason over log semantics, applying them to every line is too costly for continuous monitoring. We present \FACET{} (\underline{F}ailure-\underline{A}ware \underline{M}ixture-of-\underline{E}xperts)\footnote{\url{https://github.com/HuanchiW95/FAME-Failure-Aware-Mixture-of-Experts}}, a label-efficient message-level mixture-of-experts framework that uses an LLM only once offline. We annotate at most $K$ labeled lines per template to derive binary normal/anomaly indicators and representative examples. The LLM proposes a partition of templates into failure domains, and a certification step validates the proposal before training. \FACET{} trains a lightweight router and domain experts that run on-premise and output anomaly predictions and failure-domain labels. On BGL, \FACET{} achieves $F_1 = 98.16$ at $K = 100$ reducing annotation effort by $76{\times}$ and detects 97.7\% of anomalies from unseen EventIDs. On Thunderbird, \FACET{} reaches $F_1 = 99.95$ with perfect recall.

\end{abstract}


%% file: introduction_final.tex
\section{Introduction}
\label{sec:intro}

Modern distributed systems generate tens of thousands of log lines per hour across heterogeneous components such as kernels, storage controllers, schedulers, and application services~\cite{logzip}. These messages record execution events, state transitions, warnings, and errors, forming the primary foundation for runtime observability. Anomalies that signal hardware faults, process crashes, or service degradation are typically rare, often accounting for less than $5\%$ of generated lines. This severe class imbalance makes fine-grained log anomaly detection a persistent operational challenge~\cite{logparsingeval}.

\begin{figure}[t]
\centering
\includegraphics[width=\linewidth]{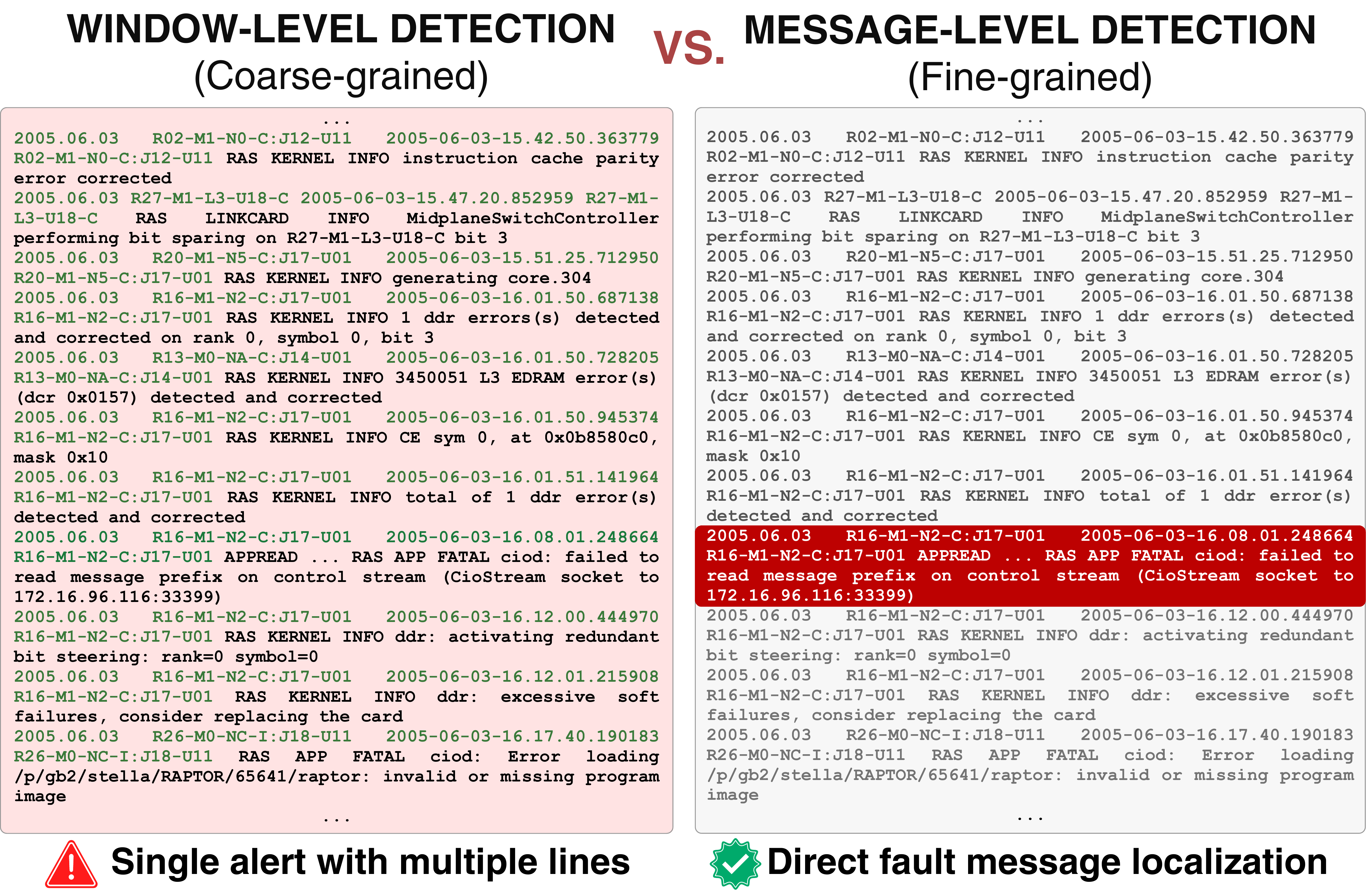}
\vspace{-5mm}

\caption{Window-level vs message-level in log anomaly detection}
\label{fig:granularity}
\end{figure}

Most deployed log anomaly detectors operate at the \emph{window-} or \emph{session-level}, assigning a single anomaly label to a contiguous group of lines delimited by a session key or a fixed time window~\cite{logbert,deeplog,loganomaly}. While computationally convenient, this granularity imposes substantial downstream cost. A window-level alert localizes suspicion to a \emph{region} rather than a \emph{specific message}, and dismissing a false alarm typically requires reviewing hundreds of routine log lines. In one production setting that motivates our work, each false-positive window forces a manual review of roughly $300$ interleaved lines, consuming on the order of $2.5$~engineer-hours before clearance. Under realistic alarm rates, the cumulative review cost of false positives can exceed the investigation budget for genuine incidents. Figure~\ref{fig:granularity} contrasts the two granularities.

More broadly, in interleaved production streams where session boundaries are weak or unavailable~\cite{li2022swisslog}, per-message output is often the only actionable granularity. We therefore target \emph{message-level} (synonymously, line-level) anomaly detection, where each raw log message is individually classified as normal or anomalous. Three properties of production log streams make this task substantially harder than its window-level counterpart.

\begin{enumerate}[leftmargin=*]
  \item \textbf{C1: Template ambiguity.} A single parser-derived event template can host both benign and anomalous instances depending on parameter values. Template identity alone cannot separate normal from anomalous lines~\cite{le2022log}.
 
  \item \textbf{C2: Heterogeneous failure modes.} Production anomalies span memory errors, filesystem faults, network partitions, and software exceptions. A single global classifier must fit one decision boundary across all these populations, trading recall against precision under severe imbalance.
 
  \item \textbf{C3: Label scarcity.} Labeling a production stream at message granularity requires reviewing millions of lines, an annotation effort that is impractical in most operational settings.
\end{enumerate}

Prior work includes both sequence/window-level detectors and message-level detectors, but the combined setting of line-level detection, heterogeneous failure modes, and limited labeling remains underexplored. Template and feature-based detectors~\cite{drain,le2022log,logif} can support line-level baselines, but they typically rely on template identity or global feature representations, which limits their ability to separate benign and anomalous instances sharing the same EventID (C1). Neural sequence detectors~\cite{deeplog,loganomaly,logbert,huang2020hitanomaly,zhang2019robust} aggregate evidence over sessions or windows, producing group-level labels that do not directly localize the responsible line (C1, C2). LLM-based detectors~\cite{loggpt,rcacopilot,cui2025aetherlog} can reason over log semantics at line level, but continuous per-line inference incurs recurring API cost, latency, and data-egress overhead at production scale (C3). Section VI develops these comparisons in detail.

We present \FACET(\underline{F}ailure-\underline{A}ware \underline{M}ixture-of-\underline{E}xperts), a framework for message-level log anomaly detection built on a \emph{failure-aware mixture-of-experts} (MoE) architecture. The core insight is that LLM reasoning is needed only once. During an offline setup phase, the LLM proposes a partition of parser-derived event template into analyst-readable groupings called \emph{failure domains}. After this one-time partitioning, the LLM is never invoked again, and all deployment-time inference runs on lightweight on-premise models.
\FACET is structured as a sparse MoE with top-1 routing~\cite{moe_original,switch}, with two departures from the standard paradigm. First, experts are \emph{semantically grounded}: each corresponds to a named failure domain proposed by the LLM, rather than emerging implicitly from end-to-end training. This decomposition addresses C2 by letting each expert fit a decision boundary over a coherent sub-population instead of the union of all failure types. Second, the architecture is \emph{asymmetric}. Under $K$-shot sampling with a rare-anomaly prior, observing all-anomalous labels in a domain is statistically conclusive, while observing all-normal labels is not (Section~\ref{sec:asym-confidence}). \FACET exploits this asymmetry: pure-anomaly domains are resolved by routing alone, while mixed domains employ independently calibrated BERT~\cite{bert} classifiers. 

To control annotation cost (C3), \FACET requires at most $K$ labeled lines per event template. At the best result $K\!=\!100$ on BGL~\cite{bgl}, this yields $53{,}287$ labeled lines, an $76\times$ reduction relative to fully labeling the $4.7$M-line offline region. On Thunderbird~\cite{loghub}, where anomalous templates are largely disjoint from normal ones, \FACET reaches $F_1\!=\!99.95$. We report Thunderbird as a closed-world sanity check, with BGL as the primary evidence for the realistic intra-template-mixed regime.

The main contributions are organized as follows:

\begin{itemize}[leftmargin=*]
  \item \textbf{Message-level detection with failure-domain routing.} \FACET produces both a binary anomaly decision and a routing label for every log line, enabling direct fault localization without session reconstruction.
 
  \item \textbf{Failure-aware asymmetric MoE.} We formalize an asymmetric confidence property of $K$-shot sampling under the rare-anomaly prior, and use it to design an MoE in which pure-anomaly domains are resolved by routing alone while mixed domains employ per-message classifiers. Expert boundaries are defined by a one-time LLM partitioning and validated deterministically against $K$-shot labels.
 
  \item \textbf{Fully on-premise inference with minimal labeling.} After a single offline LLM invocation, all inference runs on local models, eliminating recurring API cost and data egress. At $K\!=\!100$, \FACET requires $76\times$ fewer labels than full annotation and reaches $F_1\!=\!98.16$ (BGL) and $F_1\!=\!99.95$ (Thunderbird), processing up to $1.20$M lines/hour on a single 4-GPU node at a one-time cost of \$$10.23$ versus \$$6{,}698$--\$$10{,}047$ per run for frontier-LLM inference. Semantic routing further enables generalization to unseen templates, detecting $97.7\%$ of anomalies from event types absent during setup.
\end{itemize}

%% file: Preliminary_final.tex
\section{Preliminaries}
\label{sec:preliminaries}

\subsection{Problem Formulation}
\label{sec:problem}

 
 

\emph{Message-level log anomaly detection} takes as input a single raw log message $x \in \mathcal{X}$ and produces a binary prediction $\hat{y} \in \{0,1\}$, where $\hat{y}=1$ indicates an anomaly. We consider this task under a $K$-shot labeling regime: during an offline setup phase, messages are parsed into EventIDs $e \in \mathcal{E}$ (e.g., via Drain~\cite{drain}), and at most $K$ labeled messages are available per EventID, together with additional unlabeled offline data. Evaluation uses standard per-message precision, recall, and $F_1$.
 
Beyond the binary decision, \FACET additionally produces a routing label $\hat{c} \in \{1,\ldots,C\}$ identifying the assigned \emph{failure domain}, a semantically coherent group of log events sharing a subsystem or failure mechanism. This routing output is a design goal of \FACET (not part of the general task definition) and is evaluated qualitatively in Section~\ref{sec:evaluation}.

\subsection{Sparse Mixture-of-Experts}
\label{sec:moe}



In a sparse mixture-of-experts (MoE) model~\cite{moe_original,switch}, a gating network routes each input to the top-$k$ (typically $k\!=\!1$) of $C$ expert sub-networks, decoupling model capacity from per-example compute. The canonical recipe trains gate and experts jointly with a load-balancing loss to prevent expert collapse.
 
A notable variant replaces the learned gate with deterministic domain identity. DEMix Layers~\cite{gururangan2022demix} assign each expert to a known text genre and route by genre label, trading adaptive gating for interpretability and eliminating the collapse problem entirely. \FACET follows this \emph{domain-grounded} lineage, treating failure domains in logs as the analogue of text genres in DEMix, but introduces two modifications.
 
\begin{enumerate}[leftmargin=*]
\item \emph{Decoupled, supervised training.} The gate and selector are trained as classifiers over failure-domain labels. Each expert is then trained independently after routing is frozen, rather than jointly with the gate.
 
\item \emph{Asymmetric routing.} Domains whose $K$-shot samples contain only anomalous evidence carry no trained classifier and are resolved by routing alone. Mixed domains each have an independently calibrated classifier. The statistical basis for this design choice is developed next.
\end{enumerate}

\subsection{Asymmetric Confidence under $K$-Shot Sampling}
\label{sec:asym-confidence}
 
Let an EventID $e$ contain $N$ lines, of which $K$ are labeled, and let $p \ll 1$ denote the prior probability that any individual line is anomalous.
 
\smallskip
\noindent\textbf{Observation (Asymmetric $K$-Shot Confidence).}
Consider two extreme outcomes of $K$-shot sampling from $e$.
 
\emph{All $K$ labels are anomalous.} The probability that $K$ independent draws from a normal-dominated population are all anomalous is $p^K$, which for any reasonable $K$ (e.g., $K \geq 10$) and $p \ll 1$ is negligible. The unlabeled remainder therefore almost surely contains no normal lines, and routing alone suffices as the detection mechanism.

\emph{All $K$ labels are normal.} Seeing K normal lines does not rule out the possibility that the EventID still contains a small number of anomalies that the sample simply did not hit. A trained classifier is needed to recover these hidden anomalies.

\smallskip
The two cases are fundamentally asymmetric. An all-anomalous $K$-shot sample is conclusive, while an all-normal sample is not. This asymmetry provides the statistical justification for \FACET's MoE design, in which pure-anomaly domains are resolved by routing while mixed domains require per-message classifiers. Section~\ref{sec:methodology} develops the full pipeline.




%% file: methodology_final.tex
\section{Methodology}

\label{sec:methodology}

\begin{figure}[t]
\centering
\includegraphics[width=\linewidth]{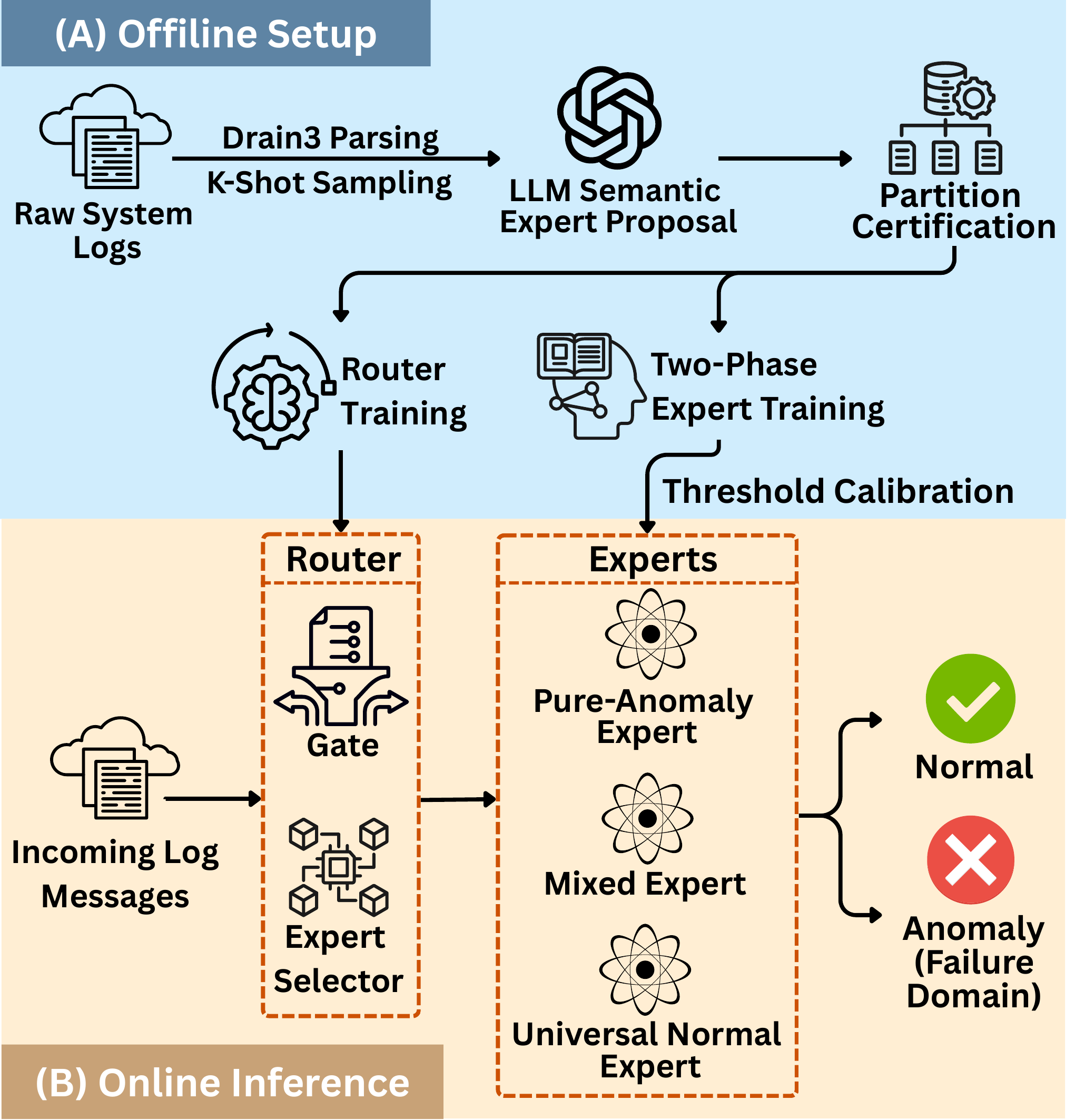}
\vspace{-5mm}
\caption{\FACET system architecture. (a)~Offline setup: raw logs are parsed by Drain3, $K$-shot labels are sampled, an LLM proposes a failure-domain partition that is then certified, and two-phase BERT experts are trained alongside a DistilBERT gate and selector. This stage is executed once. (b)~Online inference: the trained router directs each incoming log line to the appropriate expert. All inference runs on-premise with no LLM dependency.}
\label{fig:arch}
\end{figure}

\FACET operates in two stages (Figure~\ref{fig:arch}). An \emph{offline setup} stage, executed once, builds the entire detection pipeline: it parses raw logs into event templates, samples $K$-shot labels, partitions templates into failure domains, trains a lightweight router and per-domain experts, and calibrates decision thresholds. An \emph{online inference} stage then routes each incoming log line through the trained pipeline with no further LLM involvement. The remainder of this section follows the offline pipeline in execution order, then describes the inference procedure.


We extend the notation of Section~\ref{sec:problem}. Let $\{(x_i,y_i,e_i)\}_{i=1}^N$ denote the chronologically ordered offline log lines, where $e_i \in \mathcal{E}$ is the Drain-parsed EventID of line $x_i$. The first $85\%$ of lines form the offline region and the final $15\%$ the held-out test set. Failure domains are indexed by $c \in \{1,\dots,C\}$, with the distinguished class $u$ denoting \texttt{UNIVERSAL\_NORMAL}. The certified partition is a deterministic mapping $\pi\!:\!\mathcal{E}\!\to\!\{1,\dots,C\}$, and each domain carries a type flag $\rho_c \in \{0,1\}$ ($\rho_c\!=\!1$ for pure-anomaly, $\rho_c\!=\!0$ for mixed). The PU-normal pool $\mathcal{I}_{\text{normal}}$ contains all offline lines not labeled anomalous in the $K$-shot sample.

\subsection{Log Parsing and $K$-Shot Sampling}
\label{sec:parsing}

Before any model is trained, raw messages must be abstracted into templates so that structural patterns become visible. We strip label tokens to prevent leakage, then apply Drain3~\cite{drain} with similarity threshold $0.5$ and tree depth $4$ (fixed across all datasets). Each message $x_i$ is mapped to an EventID $e_i$ that replaces instance-specific tokens with wildcards while preserving structural semantics.
 
To control annotation cost (C3), we limit labeling to at most $K$ lines per EventID, deterministically selecting the $K$ most recent labeled lines in the offline region to reduce temporal mismatch with deployment. These samples are split chronologically 80/20 per EventID into a training subset (used in Phase~2 fine-tuning, Section~\ref{sec:experts}) and a calibration subset (reserved for threshold calibration, Section~\ref{sec:calibration}). From each EventID's sample, we derive two binary signals, \textsc{HasNormalSignal} and \textsc{HasAnomalySignal}, and extract at most one representative normal and one anomalous line. These signals and representatives are the only inputs to the LLM in the next stage.

\subsection{Failure-Domain Partitioning}
\label{sec:partitioning}

A single global classifier over the full log vocabulary faces the heterogeneous-failure-mode problem (C2). \FACET addresses this by decomposing the log space into \emph{failure domains}, which are partitions of EventIDs targeting narrower log regions so that each expert fits a tighter decision boundary. The partition can be proposed by an LLM to produce analyst-readable group names, or derived automatically (e.g., TF-IDF K-means).

\subsubsection{LLM-Guided Semantic Expert Proposal}
\label{sec:llm_partition}

The LLM receives a per-EventID statistics table containing the total offline line count, the binary signals, and the representative lines described above. A structured prompt instructs the LLM to group EventIDs by semantic similarity of log content, using failure mechanism as an interpretive guide, and ignoring boilerplate tokens. The LLM determines the number of groups $C$ and outputs a partition $\mathcal{E} = \bigcup_{c=1}^C \mathcal{E}_c$ that must include one group named \texttt{UNIVERSAL\_NORMAL}. Any unassigned EventID defaults to \texttt{UNIVERSAL\_NORMAL}. These groupings are structural priors for expert decomposition, not independently validated failure taxonomies.

\subsubsection{Feasibility Validation and Partition Certification}
\label{sec:feasibility}

Because the LLM output is non-deterministic, we treat it as a \emph{proposal} and subject it to deterministic feasibility validation using the K-shot binary signals.
For each failure-domain class $c \neq u$:
\begin{enumerate}[label=(\roman*)]
    \item \textbf{Pure-anomaly expert} ($\rho_c=1$). All EventIDs in $\mathcal{E}_c$ satisfy \textsc{HasAnomalySignal}$=1$ and \textsc{HasNormalSignal}$=0$. Routing itself serves as the detector and no expert model is trained. We require the group's sample logs to be sufficiently distinct from the \texttt{UNIVERSAL\_NORMAL} (TF-IDF cosine similarity $< 0.3$); otherwise the group is dissolved into \texttt{UNIVERSAL\_NORMAL}.
    \item \textbf{Mixed expert} ($\rho_c=0$). At least one EventID in $\mathcal{E}_c$ has \textsc{HasNormalSignal}$=1$. Both Phase~1 and Phase~2 training are applied.
    \item \textbf{No anomaly signal}. If all EventIDs in $\mathcal{E}_c$ satisfy \textsc{HasAnomalySignal}$=0$, the group is dissolved into \texttt{UNIVERSAL\_NORMAL}.
\end{enumerate}
After certification, the partition is frozen as the mapping $\pi$ together with the type flags $\{\rho_c\}$. Condition~(i) prevents pure-anomaly experts that are indistinguishable from normal logs; condition~(iii) prevents experts with no positive training signal. In both cases dissolved EventIDs fall back to \texttt{UNIVERSAL\_NORMAL}, so the system is designed to degrade gracefully on suboptimal partitions; see Section~\ref{sec:ablation} for the empirical lower bound.

\subsection{Router Training}
\label{sec:router}

With the partition certified, the next task is to learn a router that directs each line to the correct domain at inference time. The router is trained before expert models so that routing quality can be verified before the more expensive expert-training stage. Because the router evaluates every log line, both of its components are instantiated as lightweight \texttt{distilbert-base-uncased} classifiers~\cite{distilbert}.


\subsubsection{Gate}

The gate $g_\phi:\mathcal{X}\rightarrow[0,1]$ is a binary DistilBERT classifier that predicts whether a line belongs to any expert domain. Its labels are $y_i^{\text{gate}}=\mathbf{1}[\pi(e_i)\neq u]$. All expert lines are retained. \texttt{UNIVERSAL\_NORMAL} lines are subsampled to a 3:1 ratio to reduce training cost. We fine-tune all DistilBERT layers with a linear head using Focal Loss with $\gamma=2$ and $\alpha=0.75$. We fix the routing threshold at $\tau_{\text{gate}}=0.5$. Training stops once validation recall at this threshold reaches $0.95$. This corresponds to fewer than $5\%$ of expert-domain lines in the router validation split being misrouted to \texttt{UNIVERSAL\_NORMAL}.

\subsubsection{Selector}

The selector $h_\psi:\mathcal{X}\rightarrow\{1,\dots,C\}\setminus\{u\}$ is a multiclass DistilBERT classifier that routes expert-domain lines to a specific expert. All expert-domain training lines are used without subsampling. To address class imbalance, class weights are defined as
\begin{equation}
w_c = \frac{N_{\text{total}}}{(C-1)\cdot \text{count}_c},
\label{eq:classweight}
\end{equation}
where $\text{count}_c$ is the number of training lines in expert $c$. Training stops when overall accuracy on the router validation split reaches $0.80$.

\subsection{Two-Phase Expert Training}
\label{sec:experts}

\texttt{UNIVERSAL\_NORMAL} serves as both the default inference path for non-expert lines and a catch-all anomaly detector for anomalies missed by the gate. Each mixed expert with $\rho_c=0$, including \texttt{UNIVERSAL\_NORMAL}, undergoes both Phase~1 and Phase~2 training. Pure-anomaly experts with $\rho_c=1$ require no expert model.

\subsubsection{Phase 1: Domain-adaptive MLM pre-training}

BERT's original pre-training corpus is general web text, which is a poor lexical match for log syntax. To adapt the language model to log vocabulary, we continue pre-training \texttt{bert-base-uncased} with masked language modeling (masking rate $15\%$)~\cite{bert}. The pre-training corpus for all experts is the \texttt{UNIVERSAL\_NORMAL} PU-normal pool $\mathcal{I}_{\text{normal}} \cap \{i\!:\!e_i\!\in\!\mathcal{E}_u\}$, capped at $200{,}000$ uniformly sampled lines.
Using a shared normal-only corpus serves two purposes. First, it avoids contaminating the language model with anomalous patterns, since the PU-normal pool excludes all K-shot-labeled anomalous lines by construction. Residual unlabeled anomalies may remain, but their effect is bounded by the low base anomaly rate typical of production logs. Second, it allows all experts to share a single Phase~1 checkpoint rather than pre-training independently.

\subsubsection{Phase 2: Supervised fine-tuning}

A binary classification head is added on top of the BERT \texttt{[CLS]} representation. Following Sun et al.~\cite{sun2019bert}, we freeze the embedding layer and the first 10 of 12 transformer layers. We train only the top two layers and the head. For each mixed expert $c$, training uses labeled anomaly lines from the 80\% K-shot training split together with normal lines drawn from the \texttt{UNIVERSAL\_NORMAL} PU-normal pool. For \texttt{UNIVERSAL\_NORMAL}, the anomaly set includes all expert EventIDs so that it can act as a fallback detector for misrouted anomalies. The normal pool is capped at $10\times$ the anomaly count for \texttt{UNIVERSAL\_NORMAL} and $20\times$ for expert experts. We optimize Focal Loss with $\gamma=2$ and $\alpha=0.75$, which emphasizes hard anomaly examples under severe class imbalance:
\begin{equation}
\mathcal{L}_{\text{focal}}(\hat{p}, y) = -\alpha_t \,(1 - \hat{p}_t)^\gamma \log \hat{p}_t,
\label{eq:focal}
\end{equation}
where $\hat{p}_t = \hat{p}$ if $y = 1$ and $1 - \hat{p}$ otherwise, and $\alpha_t = \alpha$ if $y = 1$ and $1 - \alpha$ otherwise. Small datasets with fewer than 4{,}000 total lines are trained for a fixed 500 gradient steps. AUROC-based early stopping is checked every 50 steps. Larger datasets use epoch-based training with validation AUROC and patience $P_2=3$. The anomaly score is
\begin{equation}
s_c(x_i)=\sigma(f_{\theta_c}(x_i))\in[0,1].
\label{eq:score}
\end{equation}

\subsection{Threshold Calibration}
\label{sec:calibration}

For each mixed expert $c$, the decision threshold $\tau_c$ is calibrated on the held-out 20\% K-shot calibration subset defined above. We select $\tau_c$ by maximizing F1 over 1{,}000 score percentiles, subject to recall $\ge 0.90$. If fewer than 1{,}000 unique scores are available, we search over all unique scores. For \texttt{UNIVERSAL\_NORMAL}, we additionally fuse the expert score $s_u(x_i)$ with the gate output $g_\phi(x_i)$:
\begin{equation}
\tilde{s}_u(x_i)=\sigma\!\left(\mathrm{logit}(s_u(x_i)) + w\cdot \mathrm{logit}(g_\phi(x_i))\right),
\label{eq:fusion}
\end{equation}
where both $w$ and the final threshold are selected by grid search on the same calibration subset. Because this subset is small in the low-$K$ regime, threshold estimates may be unstable. We revisit this effect in Section~\ref{sec:evaluation}.

\subsection{Routing and Inference}
\label{sec:inference}

At inference time, each log line follows one of three paths:
\begin{equation}
\hat{y}_i=
\begin{cases}
\mathbf{1}[\tilde{s}_u(x_i)\ge\tau_u] & \text{if } g_\phi(x_i)<0.5,\\
1 & \text{if } g_\phi(x_i)\ge0.5 \text{ and } \rho_{c^\ast}=1,\\
\mathbf{1}[s_{c^\ast}(x_i)\ge\tau_{c^\ast}] & \text{if } g_\phi(x_i)\ge0.5 \text{ and } \rho_{c^\ast}=0,
\end{cases}
\label{eq:inference}
\end{equation}
where $c^\ast=h_\psi(x_i)$ is the selector's chosen domain. Lines routed to \texttt{UNIVERSAL\_NORMAL} (gate score below $0.5$) are scored by the fused model $\tilde{s}_u$ and flagged without a failure-domain label. Lines routed to a pure-anomaly domain are flagged immediately. Lines routed to a mixed domain are scored by the domain's expert against its calibrated threshold. In the latter two cases, the routing label $\hat{c}_i\!=\!c^\ast$ is emitted alongside the binary decision, providing a failure-domain annotation for downstream triage.

%% file: evaluation_final.tex
\section{Evaluation}
\label{sec:evaluation}

We evaluate \FACET on two widely used system log benchmarks to answer six research questions:

\begin{enumerate}[label=\textbf{RQ\arabic*.},leftmargin=*,topsep=2pt,itemsep=0pt]
  \item Does K-shot expert routing outperform single-model and classical baselines?
  \item How does \FACET compare to direct LLM inference in detection performance and deployment cost?
  \item How does detection performance vary with the annotation budget $K$?
  \item How sensitive is \FACET to the choice of LLM for semantic grouping?
  \item What is the contribution of each design component?
  \item How reliably does \FACET route anomalies to their correct failure domain, and how does it handle unseen EventIDs?
\end{enumerate}

\subsection{Evaluation Protocol}
All methods operate strictly at the message-level, without session aggregation or temporal context. Logs are split chronologically into an offline region and a held-out test region. The first 85\% of lines form the offline region and the final 15\% form the test set. Within the offline region, K-shot labeled lines are sampled per EventID and further split chronologically into a Phase~2 training subset and a held-out calibration subset. We report Precision, Recall, F1, and AUROC as percentages on a 0--100 scale for readability; for example, 98.16 corresponds to 0.9816 in the standard [0,1] representation. Two decimal places are retained throughout because, at the scale of our test sets, a difference of 0.01 in reported F1 corresponds to roughly 70 individual lines, so small changes in false positives or false negatives remain meaningful and are not rounding noise. \FACET and its ablations use the calibration protocol described in Section~\ref{sec:methodology}. External score-based baselines are calibrated on the same held-out subset using validation-based threshold selection.

\begin{table*}[t]
\centering
\renewcommand{\arraystretch}{0.96}
\caption{Message-level anomaly detection at $K{=}100$ on BGL and Thunderbird. LLM baselines use many-shot in-context anomaly-only demonstrations. AUROC is omitted as LLMs produce hard binary predictions. Results report the best of 5 independent seeds for baselines, ablations and \FACET; run-to-run variance is analyzed in Section~\ref{sec:k_sensitivity}.
}
\label{tab:main_results}
\begin{tabular}{lcccc|cccc}
\toprule
& \multicolumn{4}{c|}{\textbf{Blue Gene/L (BGL)}} & \multicolumn{4}{c}{\textbf{Thunderbird}} \\
\textbf{Method} & \textbf{Prec.} & \textbf{Rec.} & \textbf{F1} & \textbf{AUROC} & \textbf{Prec.} & \textbf{Rec.} & \textbf{F1} & \textbf{AUROC} \\
\midrule
\multicolumn{9}{l}{\textit{Baselines}} \\
Drain+RForest              & 99.91 & 20.19 & 33.59 & 60.14 & 100.00 & 100.00 & 100.00 & 100.00 \\
TF-IDF+IForest             & 17.13 & 71.84 & 27.67 & 81.14 & 38.48 & 52.99 & 44.58 & 93.79 \\
SBERT+LR                   & 49.89 & 96.09 & 65.68 & 96.64 & 73.48 & 91.95 & 81.68 & 99.32 \\
LogMoE                    & 7.14 & 100.00 & 13.33 & 46.88 & 7.69 & 100.00 & 14.27 & 55.97 \\
\midrule
\multicolumn{9}{l}{\textit{FAME without LLM grouping}} \\
\FACET w/ TF-IDF grouping   & 97.63 & 88.80 & 93.01 & 99.97 & 97.41 & 100.00 & 98.69 & 100.00 \\
\midrule
\multicolumn{9}{l}{\textit{Direct LLM Inference (in-context)}} \\
GPT-5.4                    & 92.32 & 100.00 & 96.01 & --      & 100.00 & 100.00 & 100.00 & --      \\
GPT-5-mini                 & 80.33 & 100.00 & 89.09 & --      & 65.48 & 100.00 & 79.14 & --      \\
Claude Sonnet 4.6          & 92.25 & 100.00 & 95.97 & --      & 84.29 & 100.00 & 91.47 & --      \\
Claude Haiku 4.5           & 94.02 & 100.00 & 96.92 & --      & 94.41 & 100.00 & 97.13 & --      \\
Gemini 3.1 Pro             & 92.32 & 100.00 & 96.01 & --      & 99.48 & 100.00 & 99.74 & --      \\
Gemini 3 Flash             & 84.52 & 100.00 & 91.61 & --      & 92.59 & 100.00 & 96.15 & --      \\
\midrule
\multicolumn{9}{l}{\textit{Ablations}} \\
Single BERT (P1+P2)                 & 17.46 & 98.24 & 29.65 & 95.39 & 11.01 & 91.83 & 19.67 & 68.65 \\
Single BERT (P2 only)               & 36.91 & 39.38 & 38.10 & 96.05 & 13.23 & 90.29 & 23.09 & 64.19 \\
Single Qwen 3.5-0.8B (P2 only)      & 7.35 & 88.36 & 13.58 & 78.84 & 7.68 & 100.00 & 14.27 & 60.43 \\
No-Gate (symmetric routing)         & 95.30 & 87.05 & 90.99 & 99.62 & 99.90 & 100.00 & 99.95 & 100.00 \\
\FACET (P2 only)                    & 96.01 & 99.77 & 97.85 & 99.92 & 99.90 & 100.00 & 99.95 & 100.00 \\
\midrule
\FACET (K=100)                      & 98.18 & 98.14 & 98.16 & 99.94 & 99.90 & 100.00 & 99.95 & 100.00 \\
\bottomrule
\end{tabular}
\end{table*}

We evaluate on BGL~\cite{bgl} and Thunderbird~\cite{loghub} datasets because both provide native line-level anomaly labels and represent two different regimes. BGL is harder because many templates mix normal and anomalous lines across heterogeneous subsystems. Thunderbird is easier because anomalous EventIDs are largely distinct from normal ones. Many widely used log benchmarks, such as HDFS, provide only sessional- or block-level labels and therefore do not support direct line-level evaluation without extra relabeling assumptions.

\paragraph{Blue Gene/L (BGL)}
The BGL dataset contains 4{,}747{,}963 log lines collected from an IBM Blue Gene/L supercomputer, of which 348{,}460 (7.3\%) are labeled anomalous. The test partition contains 712{,}195 lines including 46{,}278 anomalies.

\paragraph{Thunderbird}
Thunderbird is a 5M-line subset of the Thunderbird supercomputer log, with 7.6\% anomalous lines. The test partition contains 750{,}000 lines including 56{,}800 anomalies.

\subsection{Baseline Methods}
\label{sec:baselines}

We compare \FACET against three non-routed line-level baselines, six 
frontier LLMs evaluated as direct classifiers, a non-LLM grouping variant 
of \FACET, and architectural ablations. All methods share the same Drain3 
parsing and chronological test partition.

\paragraph{Baselines}
Drain+Random Forest (RForest) is a supervised EventID-only baseline using 
one-hot Drain3 EventIDs and a Random Forest trained on the same $K$-shot 
labeled lines as \FACET. SBERT+Logistic Regression (LR) uses frozen 
Sentence-BERT embeddings~\cite{reimers2019sentence} trained on the same 
$K$-shot labeled lines, providing a strong global semantic baseline. 
TF-IDF+Isolation Forest (IForest) is an unsupervised baseline trained on 
TF-IDF token bigram features over the PU-normal pool. LogMoE is a 
two-expert BERT+LoRA mixture-of-experts baseline adapted for line-level 
scoring, with a soft-weighted source-system gate trained on BGL and 
Thunderbird $K$-shot data jointly~\cite{11334514}.

\paragraph{\FACET without LLM Grouping}
\FACET w/ TF-IDF grouping is a non-LLM variant of \FACET in which the 
offline expert partition is produced by TF-IDF similarity rather than 
by an LLM.

\paragraph{Direct LLM Inference}
We evaluate six frontier LLMs as direct log-line classifiers, representing 
the alternative of always-on LLM inference without a trained detection 
model. For each of three families, we select one high-capability and one 
cost-efficient model: GPT-5.4 with GPT-5-mini~\cite{openai_gpt5_2026}, 
Claude Sonnet 4.6 with Claude Haiku 4.5~\cite{anthropic_claude_2026}, and 
Gemini 3.1 Pro with Gemini 3 Flash~\cite{google_gemini_2026}. Each log 
line is submitted individually with a structured prompt asking for a binary 
decision. The LLM receives at least one anomalous sample per EventID as 
in-context examples. No fine-tuning is performed.

\paragraph{Ablations}
Single BERT (P1+P2) is trained with the same two-phase protocol as \FACET 
but without routing or specialization, isolating the contribution of expert 
routing. Single BERT (P2 only) disables Phase~1 MLM pre-training, isolating 
the contribution of domain-adaptive initialization. Both are instantiated as 
\texttt{bert-base-uncased}. Single Qwen~3.5-0.8B (P2 only) replaces BERT 
with Qwen~3.5-0.8B fine-tuned with LoRA under the same Phase~2 objective, 
testing whether the weakness of a single global detector is 
architecture-specific. No-Gate removes the asymmetric gate, using symmetric 
routing only. \FACET (P2 only) removes Phase~1 from the full pipeline while 
retaining all other components.

\subsection{Single-Model and Classical Baselines (RQ1)}
\label{sec:rq1}
Table~\ref{tab:main_results} shows different behavior on the two datasets. On BGL, \FACET clearly outperforms all non-routed baselines. Drain+RForest reaches F1\,=\,33.59. TF-IDF+IForest reaches 27.67. SBERT+LR reaches 65.68 and has AUROC\,=\,96.64. This means that a strong global semantic representation can rank anomalies well, but still struggles to place one global decision boundary over many different anomaly types. \FACET improves on this by splitting the log space into smaller expert domains.

Thunderbird is an easier dataset as a closed-world control. Drain+RForest already reaches perfect performance, which means anomalous templates are largely separable from normal ones. SBERT+LR is also strong with F1\,=\,81.68. IForest is weaker, which suggests that simple unsupervised scoring is not enough even when the dataset is easier. \FACET should therefore not be read as uniquely solving a hard detection problem on Thunderbird. Instead, it matches near-perfect performance on an easier benchmark while yielding much larger gains on BGL.

The single-model neural ablations remain far below \FACET on both datasets. On BGL, this supports the value of expert routing. On Thunderbird, the contrast with Drain+RForest shows that the main issue is not lack of signal in the data. The main issue is learning one well-calibrated global detector under strong class imbalance.

We compare \FACET against LogMoE, which is baseline that shares the MoE architecture. Under the same $K{=}100$ protocol, LogMoE reaches $F_1{=}13.33$ on BGL and $F_1{=}14.27$ on Thunderbird. Both results have recall $=100.00$ but precision below $8\%$, meaning LogMoE flags every log line as anomalous. This happens because LogMoE's gate is designed for inter-system routing at the window level, not for intra-system detection within a single deployment. The large gap between LogMoE and \FACET ($F_1{=}98.16$ on BGL) shows that the gain comes from failure-domain specialization and asymmetric routing, not from MoE architecture alone.

\begin{figure*}[t]
\centering
\includegraphics[width=0.9\textwidth]{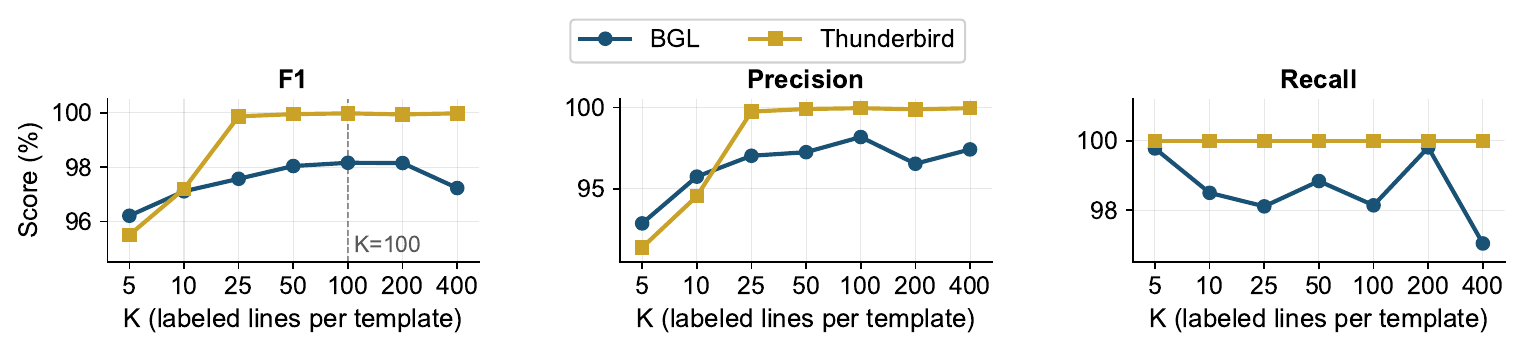}
\vspace{-5mm}
\caption{K-sensitivity on BGL and Thunderbird with best result at each K.}
\label{fig:k_sensitivity_combined}
\end{figure*}

\begin{figure*}[t]
\centering
\includegraphics[width=0.9\textwidth]{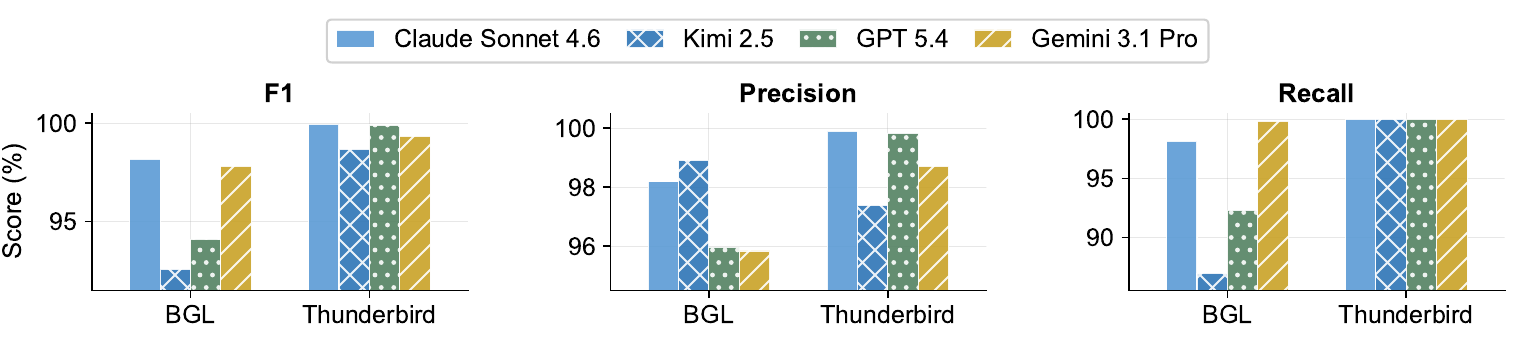}
\vspace{-5mm}
\caption{LLM grouping sensitivity on BGL and Thunderbird ($K{=}100$). Each subplot shows F1, Precision, or Recall for all four LLMs.}
\label{fig:llm_combined}
\end{figure*}
\subsection{Direct LLM Inference and Deployment Cost (RQ2)}
\label{sec:rq2}
For direct many-shot in-context LLM inference, we cannot simply include the full training set as labeled example lines due to context length, as the examples alone would exceed 2 million tokens. Instead, we construct the prompt to preserve anomaly coverage while keeping the context compact. We include at least one anomalous example from every EventID that contains anomalies, so that the LLM sees the full range of anomalous EventID types present in the offline region. When the number of anomaly-containing EventIDs is smaller than 50, we sample additional anomalous labeled lines until reaching at least 50 demonstrations for a fair comparison across datasets and models. We also evaluated normal-only prompts and mixed prompts containing both normal and anomalous examples. Normal-only prompting performed substantially worse. Mixed prompting performed similarly to anomaly-only prompting but required a larger context. We therefore use anomaly-only demonstrations as the strongest compact baseline. This yields prompt sizes of 4{,}652 tokens/call for BGL, with 96 example lines, and 2{,}367 tokens/call for Thunderbird, with 50 example lines. These counts include a 500-token system prompt, labeled demonstrations, and the test line. For cost estimation, we assume a 10-token output per call.

Table~\ref{tab:main_results} shows that on BGL, direct many-shot in-context LLM inference is strong but still below \FACET in F1. All LLM baselines achieve perfect recall at the cost of lower precision, suggesting they flag most lines as anomalous rather than learning a well-calibrated decision boundary. \FACET achieves a better precision-recall balance, reaching F1\,=\,98.16 with both high recall (98.14) and high precision (98.18). Among the evaluated models, Claude Haiku 4.5 performs best with F1\,=\,96.92, followed by GPT-5.4 and Gemini 3.1 Pro at F1\,=\,96.01. Lower-cost variants are less consistent, with GPT-5-mini achieving F1\,=\,89.09 and Gemini 3 Flash reaching F1\,=\,91.61. On Thunderbird, several frontier LLMs achieve near-perfect or perfect performance, consistent with the structural separability of that dataset.

Table~\ref{tab:llm_cost} shows a study on the corresponding cost of in-context LLM inference at standard API pricing~\cite{openai_pricing,anthropic_pricing,google_pricing}. A single full BGL test-set run costs \$10{,}047 with Claude Sonnet 4.6, \$8{,}373 with GPT-5.4, and \$6{,}698 with Gemini 3.1 Pro. Even lower-cost variants still require \$837 to \$1{,}675 per run. On Thunderbird, costs range from \$5{,}438 for Claude Sonnet 4.6 and \$4{,}532 for GPT-5.4 down to \$453 for GPT-5-mini. These API charges recur for every evaluation and scale linearly with log volume. By contrast, \FACET incurs a one-time setup cost and inference cost of \$10.23, covering 7\,h of offline training, a single LLM grouping call (\$0.05), and 0.6\,h of inference on a GCP instance with 4 T4 GPUs, 8\,vCPUs, and 52\,GB RAM at enterprise pricing (\$1.34/hr).

\begin{table}[t]
\centering
\caption{Estimated cost of many-shot in-context LLM inference on each test set at standard API pricing, compared with \FACET's one-time setup and inference cost.}
\label{tab:llm_cost}
\resizebox{\columnwidth}{!}{%
\begin{tabular}{lrr}
\toprule
\textbf{Model} & \textbf{BGL (712{,}195 lines)} & \textbf{Thunderbird (750{,}000 lines)} \\
\midrule
Claude Sonnet 4.6   & \$10{,}047 & \$5{,}438 \\
GPT-5.4             & \$8{,}373  & \$4{,}532 \\
Gemini 3.1 Pro      & \$6{,}698  & \$3{,}626 \\
Claude Haiku 4.5    & \$3{,}349  & \$1{,}813 \\
Gemini 3 Flash      & \$1{,}675  & \$906     \\
GPT-5-mini          & \$837      & \$453     \\
\midrule
\textbf{\FACET} & \multicolumn{2}{c}{\textbf{\$10.23}} \\
\bottomrule
\end{tabular}}
\end{table}

\begin{table}[t]
\centering
\caption{Annotation cost per $K$ vs.\ theoretical upper bound at full line-level labeling.}
\label{tab:labeling_cost}
\begin{tabular}{rrrrrr}
\toprule
$K$ & \multicolumn{2}{c}{\textbf{BGL}} & & \multicolumn{2}{c}{\textbf{Thunderbird}} \\
\cmidrule{2-3} \cmidrule{5-6}
  & Labels & Reduction & & Labels & Reduction \\
\midrule
 5   &   6,554 & 616$\times$ & &  3,714 & 1144$\times$ \\
10   &  11,420 & 353$\times$ & &  6,213 &  684$\times$ \\
25   &  22,711 & 178$\times$ & & 11,885 &  358$\times$ \\
50   &  34,799 & 116$\times$ & & 17,636 &  241$\times$ \\
100  &  53,287 &  76$\times$ & & 23,503 &  181$\times$ \\
200  &  85,900 &  47$\times$ & & 31,276 &  136$\times$ \\
400  & 142,690 &  28$\times$ & & 44,539 &   95$\times$ \\
\bottomrule
\end{tabular}
\end{table}

\subsection{K-Sensitivity Analysis (RQ3)}
\label{sec:k_sensitivity}

Figures~\ref{fig:k_sensitivity_combined} show F1, Precision, and Recall as a function of $K$. On BGL, the best F1 is achieved at $K{=}100$. The trend is non-monotone across budgets. At very small $K$, the calibration subset contains too few anomalies for reliable threshold selection, making the precision--recall tradeoff unstable. At $K{=}100$, \FACET reaches its strongest balance of precision and recall, and performance at $K{=}200$ is essentially equivalent. A modest decline at $K{=}400$ is driven by a precision drop with recall maintained. This pattern is consistent with expert overfitting at high $K$. Because K-shot sampling selects the most recent labeled lines per EventID, a larger $K$ draws from a wider temporal window that may shift the learned anomaly distribution away from the test region. We therefore recommend $K{=}100$ as the practical operating point, where annotation efficiency and detection reliability are jointly optimized. For annotation-constrained deployments, $K{=}25$ requires only 22,711 labeled lines on BGL (a $178{\times}$ reduction) and achieves F1\,=\,97.57, offering a practical alternative when labeling resources are limited.

On Thunderbird, all ablations remain below \FACET, consistent with the Thunderbird discussion in Section~\ref{sec:rq1} showing the dataset is structurally separable.

Table~\ref{tab:labeling_cost} reports labeled-line counts per $K$. The reduction factor is computed relative to exhaustive offline annotation as a theoretical upper bound.

We completed a run-to-run standard deviation analysis across 5 independent trials per $K$, each using a different random seed for model initialization and K-shot sampling. Standard deviation in F1 remains below 3.50 across all budgets, confirming that \FACET's performance is stable rather than dependent on a favourable random initialization. The highest variability occurs at $K{=}50$ (F1 std\,=\,3.50, Recall std\,=\,5.31), consistent with the non-monotone behaviour at that budget level where calibration is most sensitive to the specific K-shot sample drawn. At larger budgets ($K{\geq}100$), variability reduces further (F1 std\,${\leq}$\,2.13), confirming that \FACET converges to a stable operating regime with sufficient labeled data. The consistently high AUROC across all trials (${\geq}$\,99.00 at every $K$ and seed) indicates that F1 variability reflects threshold calibration sensitivity rather than learning instability. The model's discriminative ability is robust, while the specific K-shot sample drawn for calibration determines the precision-recall operating point. At low $K$, this sensitivity is amplified because the small calibration subset makes threshold selection more dependent on which anomaly examples are sampled. At high $K$, it diminishes as the calibration pool grows more representative.

\subsection{LLM Grouping Sensitivity (RQ4)}
\label{sec:llm_sensitivity}

Figure~\ref{fig:llm_combined} compares four LLMs for the offline semantic grouping step under identical pipeline settings at $K{=}100$, covering three distinct capability and deployment profiles. GPT-5.4, Claude Sonnet 4.6, and Gemini 3.1 Pro represent the high-capability closed-model tier. Kimi 2.5 represents a competitive open-weights alternative.

On BGL, Claude Sonnet 4.6 gives the strongest result among the tested grouping configurations, with F1 of 98.16, followed by Gemini 3.1 Pro at 97.78 and GPT-5.4 at 94.10. Kimi 2.5 performs lowest at 92.55. These differences reflect different grouping strategies. Gemini 3.1 Pro produces only 4 experts and reaches near-perfect recall (Rec\,=\,99.79), but generates 2{,}002 false positives. GPT-5.4 produces 8 experts and achieves high precision (95.97), but misses 3{,}563 anomalies. Kimi 2.5 is the most conservative grouping, with only 457 false positives but 6{,}047 missed anomalies. This means that coarser partitions tend to favor recall, while more conservative partitions may improve precision at the cost of missed anomalies. More broadly, these results suggest that multiple valid partitions exist. In practice, the grouping stage should be viewed as an offline design choice where different reasonable partitions can be proposed by different LLMs or repeated runs, and \FACET remains effective across this variation.

On Thunderbird, all four LLMs produce 6-expert partitions covering the same five failure categories, with only minor F1 differences (std=0.50) from boundary decisions. Grouping sensitivity is thus much lower on structurally separable datasets.

The non-LLM \FACET variant with TF-IDF grouping in Table~\ref{tab:main_results} reinforces this conclusion: the TF-IDF baseline remains strong on both datasets, reaching $F_1=93.01$ on BGL and $98.69$ on Thunderbird. This indicates that \FACET does not rely on LLM-based grouping as a core mechanism. Instead, the principal structural gain comes from expert decomposition, while LLM-based grouping serves as an offline quality booster on more heterogeneous datasets and yields semantically grounded failure domains.

Table~\ref{tab:llm_groups} details the expert domains proposed by each LLM for BGL at $K{=}100$. Despite differences in naming conventions, all LLMs identify the same core failure categories. The main structural difference is expert granularity. Claude Sonnet 4.6 and Kimi 2.5 separate some failure types more finely, while Gemini 3.1 Pro merges them into broader groups.

\begin{table}[t]
\centering
\caption{Expert domains proposed by each LLM for BGL at $K{=}100$. P = pure-anomaly (gate = detector) and M = mixed; UNIVERSAL\_NORMAL is omitted.}
\label{tab:llm_groups}
\resizebox{0.9\columnwidth}{!}{%
\begin{tabular}{llcc}
\toprule
\textbf{LLM} & \textbf{Specialist Domain} & \textbf{T} & \textbf{F1} \\
\midrule
Claude 4.6  & HARDWARE\_MACHINE\_CHECK\_INTERRUPT    & M & \multirow{9}{*}{98.16} \\
            & DDR\_MEMORY\_ERROR                     & M & \\
            & KERNEL\_PANIC\_AND\_FATAL\_TERMINATION & P & \\
            & CIOD\_IO\_DAEMON\_ERROR                & P & \\
            & LUSTRE\_FILESYSTEM\_MOUNT\_ERROR       & P & \\
            & NETWORK\_LINK\_AND\_SWITCH\_FAILURE    & P & \\
            & PROGRAM\_INTERRUPT\_AND\_FP\_EXCEPTION & M & \\
            & BGLMASTER\_CONTROL\_DAEMON\_FAILURE    & M & \\
            & POWER\_AND\_HARDWARE\_DEACTIVATION     & P & \\
\midrule
Kimi 2.5    & MEMORY\_HARDWARE\_FAILURE              & M & \multirow{7}{*}{92.55} \\
            & KERNEL\_FATAL\_ERROR                   & P & \\
            & FILESYSTEM\_MOUNT\_FAILURE             & P & \\
            & NETWORK\_COMMUNICATION\_ERROR          & P & \\
            & MIDPLANE\_SWITCH\_FAILURE              & P & \\
            & CONTROL\_SYSTEM\_FAILURE               & M & \\
            & PROGRAM\_INTERRUPT\_EXCEPTION          & M & \\
\midrule
GPT 5.4     & HARDWARE\_MACHINE\_CHECK\_ERRORS       & M & \multirow{8}{*}{94.10} \\
            & INTERRUPT\_AND\_EXCEPTION\_FAULTS      & M & \\
            & FILESYSTEM\_AND\_STORAGE\_FAILURES     & P & \\
            & CIOD\_MAILBOX\_COMMUNICATION\_ERRORS   & P & \\
            & RTS\_KERNEL\_PANIC\_AND\_ASSERTIONS    & P & \\
            & NETWORK\_FABRIC\_ERRORS                & P & \\
            & POWER\_AND\_SWITCH\_HARDWARE\_FAULTS   & P & \\
            & SYSTEM\_SERVICE\_RUNTIME\_CRASHES      & M & \\
\midrule
Gemini 3.1 Pro & HARDWARE\_MEMORY\_FAILURE          & M & \multirow{4}{*}{97.78} \\
               & NETWORK\_COMMUNICATION\_FAILURE     & P & \\
               & FILESYSTEM\_IO\_FAILURE             & P & \\
               & KERNEL\_SOFTWARE\_FAILURE           & M & \\
\bottomrule
\end{tabular}}
\end{table}

\subsection{Ablation Study (RQ5)}
\label{sec:ablation}

In the ablation study, all variants use the same calibration procedure as the full \FACET pipeline in Section~\ref{sec:calibration} to ensure a fair comparison. At $K{=}100$ on BGL, all single-model ablations perform far worse than \FACET. Single BERT (Phase~1+2) reaches only $F_1=29.65$. Removing Phase~1 increases this to $38.10$, but recall drops to $39.38$. Single Qwen~3.5-0.8B performs worse still under the same calibration setting, reaching $F_1=13.58$. These results indicate that a single global detector does not produce anomaly scores strong enough to meet \FACET's recall target while maintaining competitive precision on BGL's heterogeneous anomaly space.

To better isolate backbone effects, we also removed the recall-floor constraint for the single-model baselines. Under this more permissive setting, Single BERT (Phase~2 only) improves to $F_1=48.91$ with recall $37.35$, and Single Qwen~3.5-0.8B improves substantially to $F_1=84.61$ with recall $75.55$. This shows that a stronger decoder-style backbone can improve the precision--recall tradeoff of a single global detector, but it still remains well below \FACET's $F_1=98.16$ while increasing inference time by $3.5\times$ relative to BERT. Overall, the dominant gain in \FACET comes from the routed expert architecture rather than from backbone choice alone.

The additional ablations support the same interpretation. Removing the gate reduces $F_1$ to $90.99$ on BGL and is driven by a recall drop from $98.14$ to $87.05$, as anomalies are more often misrouted as normal logs under symmetric selection in the presence of extreme class imbalance. This confirms that asymmetric routing, in which the gate pre-filters normal-domain traffic before the selector, is an important design choice in the routing architecture. \FACET (Phase~2 only) removes Phase~1 from the full pipeline while retaining all other components. On BGL, this reduces $F_1$ from $98.16$ to $97.85$, a small drop driven primarily by a precision decrease ($98.18 \to 96.01$), while recall improves slightly ($98.14 \to 99.77$). This shows that Phase~1 MLM acts mainly as a precision booster within \FACET. It gives each expert a tighter normal-domain baseline that reduces false positives, while the routing architecture retains strong recall without it. On Thunderbird, removing Phase~1 has no measurable effect, with $F_1=99.95$ in both cases. 

On Thunderbird, the ablations also remain far below \FACET, but this result should be interpreted together with the perfect Drain+RForest score. The main lesson is not that Thunderbird is intrinsically difficult. Rather, global neural detectors remain poorly calibrated to this low-label setting, whereas \FACET is better aligned with the separable structure of the dataset. 

Finally, we note that in both evaluations, the partition certification step (Section~\ref{sec:feasibility}) did not dissolve any expert group proposed by the LLM. All proposed partitions passed the K-shot consistency checks without triggering the fallback mechanism, indicating that the certification step primarily serves as a safeguard against non-deterministic LLM behavior rather than as a correction mechanism that is frequently invoked. In the event that poor LLM output triggers widespread dissolution, performance would degrade toward the TF-IDF grouping variant ($F_1=93.01$ on BGL), which represents the no-LLM-guidance lower bound.

\subsection{Domain Routing Analysis (RQ6)}
\label{sec:routing_analysis}

We examine RQ6 by applying Drain3 post-inference to the BGL and Thunderbird test partitions. An EventID is the template identifier Drain3 assigns to each structurally distinct log message. This categorization does not affect detection, as \FACET routes each line using only raw log content without consulting EventIDs. On BGL, 79.8\% of anomalous test lines originate from EventIDs not seen during offline setup, indicating substantial temporal concept drift. \FACET's gate and selector generalize to these unseen templates because routing operates on semantic content rather than template identity. A previously unseen template is embedded near known templates sharing its failure vocabulary and is assigned to the semantically nearest failure domain, with only 0.02\% routing to the failsafe \texttt{UNIVERSAL\_NORMAL}. Of the 36{,}936 unknown-EventID anomalies, 97.7\% are successfully detected. This result is identical across all 5 independent seeds at $K{=}100$, as unseen templates predominantly route to pure-anomaly experts where detection follows directly from routing and is independent of seed. Because these anomalies constitute 79.8\% of all anomalous test lines, the result is not an artifact of a favorable K-shot draw. For the 20.2\% of anomalous lines whose EventIDs were seen during offline setup, all routing assignments match the LLM-proposed failure domain exactly. On Thunderbird, all 56{,}800 test anomalies match known EventIDs, achieving a domain-label precision of 100.00.

\subsection{Calibration Robustness}
\label{sec:threshold_sensitivity}

The gate recall target and the per-expert calibration floor both reflect an operational asymmetry where missed anomalies are costlier than false positives, and can be tuned for different deployment cost trade-offs. The gate recall target of $0.95$ serves as an early stopping criterion that controls this trade-off during training. Both datasets achieved recall $=0.999$ in the first epoch, so it had minimal impact here but remains a useful control parameter for deployments where longer training risks overfitting. Sweeping the per-expert calibration floor across $\{0.70,0.80,0.90,0.95\}$ yields $F_1=98.16$ in all cases, as the floor is non-binding: every expert's unconstrained threshold already achieves recall $\geq 0.95$, so all four values select the same operating point.

\subsection{Inference Efficiency}
\label{sec:efficiency}

We documented per-line inference latency over 35 runs, with 5 runs for each $K \in \{5,10,25,50,100,200,400\}$. On the same GCP node described in Section~\ref{sec:rq2}, \FACET averages 1.17M lines/hour on BGL and 1.20M lines/hour on Thunderbird. Variance across runs is negligible, indicating that inference cost is effectively independent of $K$ once the pipeline is trained.

%% file: case_study_final.tex
\section{Case Study}
\label{sec:casestudy}

\begin{figure}[t]
\centering
\includegraphics[width=0.95\columnwidth]{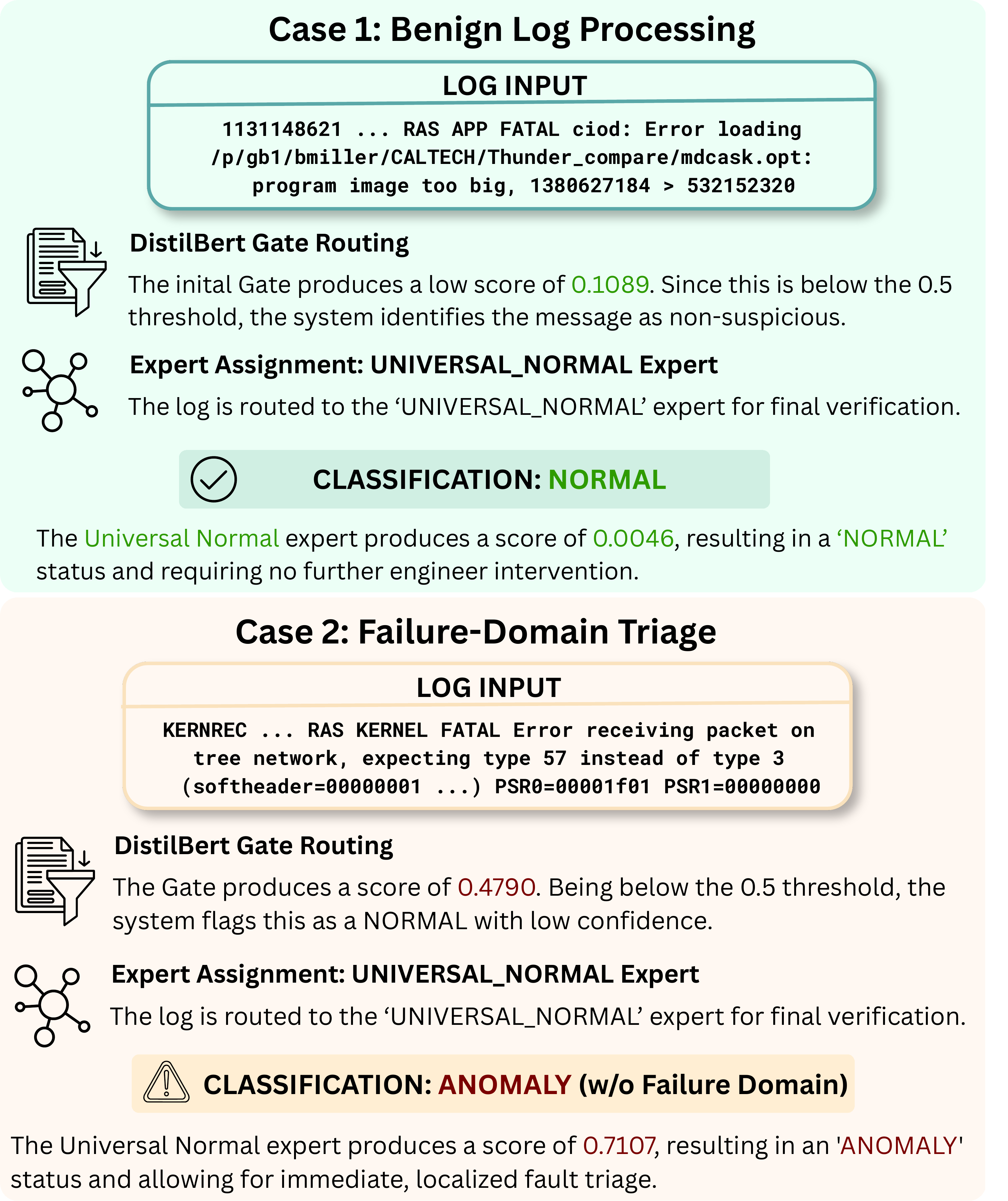}
\vspace{-3mm}
\caption{Case study with ambiguous keyword 'FATAL'}
\label{fig:casestudy}
\end{figure}

We present two inference-time edge cases to illustrate how \FACET handles ambiguous log messages, as shown in Figure~\ref{fig:casestudy}. Both logs contain the keyword \texttt{FATAL}, but they correspond to different outcomes. The first is normal and the second is anomalous.

Although the message contains \texttt{FATAL} in the first case, it indicates that the job's program binary exceeded the allowed size, which is an expected condition rather than a system failure. Both the gate and \texttt{UNIVERSAL\_NORMAL} assign high confidence to the normal class. This shows that \FACET does not rely on severity keywords alone.

In the second case, the log is a true anomaly and its template appears only in the test set. The gate output $g_\phi(x_i)=0.4790$ reflects uncertainty between a failure-domain expert and \texttt{UNIVERSAL\_NORMAL}. Despite this uncertainty, \texttt{UNIVERSAL\_NORMAL} assigns a high anomaly score of 0.7107 and correctly detects the anomaly. This example shows that \texttt{UNIVERSAL\_NORMAL} provides effective fallback behavior for unseen anomalous templates. Since the anomaly is detected by \texttt{UNIVERSAL\_NORMAL}, \FACET flags it as anomalous without assigning a failure-domain label. 

%% file: background_final.tex
\section{Related Work}
\label{sec:related}

\noindent\textbf{Log parsing and classical detectors.} Most pipelines parse raw messages into event templates that abstract instance-specific tokens. Drain~\cite{drain} is the dominant online parser. Complementary parsers include IPLoM~\cite{makanju2009clustering}, Spell~\cite{du2016spell}, and Logram~\cite{dai2020logram}, while He~et~al.~\cite{he2021survey} survey the space. Classical detectors then apply statistical models to template counts or token features, including PCA~\cite{pca_logs}, Isolation Forest~\cite{logif}, SVM and clustering~\cite{le2022log}. However, they treat all instances of a template as equivalent, which may limit their ability to distinguish normal and anomalous lines that share an EventID.

\noindent\textbf{Message-level log anomaly detection.} Although many log anomaly detectors operate at the window level, message-level detection has also been studied on datasets with native line-level labels, such as BGL and Thunderbird~\cite{bgl,loghub}. Existing line-level baselines rely on parser-derived EventIDs or templates~\cite{drain},
single global classifiers over parsed or semantic features~\cite{le2022log},
unsupervised log-message scoring~\cite{logif},
or direct LLM prompting~\cite{loggpt}. These methods provide important comparisons, but they generally treat line-level detection as a single global decision problem. \FACET instead formulates it as failure-aware routing, assigning each line to a specialized failure-domain expert and resolving pure-anomaly domains by routing alone under limited $K$-shot supervision.

\noindent\textbf{Neural sequence models and the LogBERT comparable.} DeepLog~\cite{deeplog}, LogAnomaly~\cite{loganomaly}, HitAnomaly~\cite{huang2020hitanomaly}, LogRobust~\cite{zhang2019robust}, AllInfoLog~\cite{xiao2022allinfolog}, SwissLog~\cite{li2022swisslog}, and PreLog~\cite{le2024prelog} model temporal structure within sessions to improve detection accuracy. LogBERT~\cite{logbert} is the most cited representative which applies masked language modeling to log sequences and emits an aggregate score over a session window. The contrast axis with \FACET is granularity. LogBERT's training objective and decision rule are session-aggregate by construction. Adapting it to per-message decisions is non-trivial because token scores within a line do not align with line boundaries, and the session aggregate cannot be back-projected to a single responsible message without an additional attribution model. \FACET targets per-message decisions directly, making the two approaches complementary rather than directly comparable.

\noindent\textbf{LLM-based analysis and the AdaptiveLog comparable.} LLMs have been applied to parsing through LUNAR~\cite{huang2025no} and CSLParser~\cite{hong2025cslparser}, to prompt-based log anomaly
classification through LogGPT~\cite{loggpt}, and to root-cause analysis through RCACopilot~\cite{rcacopilot}, AetherLog~\cite{cui2025aetherlog}, LogKG~\cite{sui2023logkg}, COMET~\cite{wang2024large}, and causal approaches~\cite{markakis2024logs}. Continuous per-line LLM processing introduces practical considerations such as API cost, latency, and data handling. AdaptiveLog~\cite{adaptivelog} is the closest hybrid in design philosophy. It routes high-confidence predictions to a small model and defers the uncertain residual to an LLM, which reduces but does not eliminate runtime LLM invocation. LasRCA~\cite{han2024potential} is a related variant. The crucial point for regulated deployments is that even a small deferred fraction still incurs per-call cost and, more importantly, exfiltrates the deferred lines to an external API. \FACET instead uses an LLM only during offline setup to construct the failure-domain partition, while all routing and detection run on-premise without runtime LLM calls.

\noindent\textbf{MoE variants, the DEMix comparable, and label-efficient learning.} Sparse MoE~\cite{switch} and the foundational adaptive-mixtures formulation~\cite{moe_original} establish gated conditional-compute architectures, with most work targeting generative modeling. DEMix Layers~\cite{gururangan2022demix} is the closest architectural analog to \FACET. It uses per-domain experts trained \emph{separately} from the routing, with hard routing based on domain identity, namely text genre in DEMix and failure domain in \FACET. Among MoE-based log detectors, LogMoE~\cite{11334514} is the closest related work. It trains lightweight experts on different source systems and combines them with a soft gate for cross-system transfer. \FACET addresses a different setting by performing message-level detection within one deployment system, organizing experts by failure domain rather than source system, and producing both an anomaly label and a failure-domain routing label. Thus, the two methods use MoE for different purposes. LogMoE supports cross-system sequence-level transfer, whereas \FACET supports intra-system failure-aware message-level detection. In addition, \FACET uses asymmetric routing so that pure-anomaly domains require no trained expert and can be detected by routing alone. On supervision, self-supervised LogBERT and DeepLog~\cite{logbert,deeplog}, pretraining-based PreLog~\cite{le2024prelog}, and pseudo-label PLELog~\cite{yang2021plelog} reduce labels through different mechanisms. \FACET instead spends a strict $K$-shot budget per template, achieving a $76{\times}$ reduction relative to full line-level annotation on BGL. To our knowledge, no prior system jointly targets per-message detection, named failure-domain routing labels, $K$-shot supervision, and fully on-premise deployment.

%% file: limitations_final.tex
\section{THREATS TO VALIDITY}
\label{sec:limitations}

\textbf{Internal validity.}
The $K$-shot sampling strategy selects the most recent labeled lines to approximate deployment conditions, but this recency bias may not reflect the full distributional range of each EventID. Threshold calibration relies on a held-out 20\% subset of K-shot data. At low K, the small number of anomalies can lead to unstable precision–recall tradeoffs.
\FACET commits to a fixed offline partition. Since inference does not use log parsing, the pipeline cannot automatically detect when the failure domain structure has changed. A human expert decides and triggers a full offline re-setup when a major structural change occurs, such as a software upgrade introducing new failure modes. Selective incremental retraining without full re-setup is left as future work. Finally, minor differences in baseline training or hyperparameter tuning may affect comparisons.


\textbf{External validity.}
Our evaluation uses BGL and Thunderbird, both drawn from supercomputer environments. Thunderbird’s structural separability may overestimate performance compared to noisier real-world systems. The framework also assumes $K$ labeled lines per template, which may be difficult under high template turnover.
For unseen EventIDs, failure-domain labels are assigned by semantic similarity and cannot be directly verified due to the lack of ground-truth failure types. In addition, different LLMs or prompts may produce different partitions, and real-world deployments may exhibit stronger concept drift than observed in our experiments.

%% file: conclusion_final.tex
\section{Conclusion}
\label{sec:conclusion}
We presented \FACET, a failure-aware mixture-of-experts framework for message-level log anomaly detection under limited line-level supervision. The main result of this work is that strong message-level detection can be achieved with a routed expert architecture that remains fully on-premise during inference. The LLM is used only once during offline setup to propose a semantic partition of EventIDs into failure domains. After setup, every routing and detection decision is made by lightweight local models, with no recurring API calls, no log egress, and no dependence on external services at deployment time.

Our results show that the main contribution is the architecture itself rather than reliance on an LLM. On BGL, \FACET reaches $F_1=98.16$ at $K=100$, while the non-LLM variant with TF-IDF grouping still achieves a strong $F_1=93.01$. This shows that expert decomposition and asymmetric routing, not LLM-based grouping, are the primary drivers of the performance gain. The results also show that \FACET remains practical under tighter labeling budgets, offering practitioners a useful tradeoff between annotation cost and detection performance while preserving the same fully local inference pipeline.

\FACET demonstrates that lightweight per-template annotation, offline semantic partitioning, and specialized SLM inference can be combined in a practical on-premise pipeline for line-level log analysis. One natural next step is online adaptation under schema drift, where new templates are absorbed into existing experts or trigger new specialists, and only the affected components are retrained. Another is to model temporal structure over expert activations, which could support causal incident graphs rather than only per-line anomaly labels. The framework also creates a natural entry point for combining logs with complementary operational signals such as system metrics, especially in cases where disagreement between the two may help expose silent failures or noisy components.

%% file: references.bib
@inproceedings{deeplog,
  author = {Du, Min and Li, Feifei and Zheng, Guineng and Srikumar, Vivek},
  title = {DeepLog: Anomaly Detection and Diagnosis from System Logs through Deep Learning},
  booktitle = {Proceedings of the ACM SIGSAC Conference on Computer and Communications Security},
  year = {2017},
  pages = {1285--1298}
}

@inproceedings{loganomaly,
  author = {Meng, Weibin and Liu, Ying and Zhu, Yichen and Zhang, Shenglin and Pei, Dan and Liu, Yuqing and Chen, Yihao and Zhang, Ruizhi and Tao, Shimin and Sun, Pei and Zhou, Rong},
  title = {LogAnomaly: Unsupervised Detection of Sequential and Quantitative Anomalies in Unstructured Logs},
  booktitle = {Proceedings of the International Joint Conference on Artificial Intelligence},
  year = {2019},
  pages = {4739--4745}
}

@INPROCEEDINGS{loggpt,
  author={Qi, Jiaxing and Huang, Shaohan and Luan, Zhongzhi and Yang, Shu and Fung, Carol and Yang, Hailong and Qian, Depei and Shang, Jing and Xiao, Zhiwen and Wu, Zhihui},
  booktitle={2023 IEEE International Conference on High Performance Computing \& Communications, Data Science \& Systems, Smart City \& Dependability in Sensor, Cloud \& Big Data Systems \& Application (HPCC/DSS/SmartCity/DependSys)}, 
  title={LogGPT: Exploring ChatGPT for Log-Based Anomaly Detection}, 
  year={2023},
  volume={},
  number={},
  pages={273-280},
  doi={10.1109/HPCC-DSS-SmartCity-DependSys60770.2023.00045}
  }

@article{distilbert,
  author = {Sanh, Victor and Debut, Lysandre and Chaumond, Julien and Wolf, Thomas},
  title = {DistilBERT, a Distilled Version of BERT: Smaller, Faster, Cheaper and Lighter},
  journal = {arXiv preprint arXiv:1910.01108},
  year = {2019}
}

@inproceedings{pca_logs,
  author = {Lakhina, Anukool and Crovella, Mark and Diot, Christophe},
  title = {Diagnosing Network-Wide Traffic Anomalies},
  booktitle = {Proceedings of ACM SIGCOMM},
  year = {2004},
  pages = {219--230}
}

@article{moe_original,
  author = {Jacobs, Robert A. and Jordan, Michael I. and Nowlan, Steven J. and Hinton, Geoffrey E.},
  title = {Adaptive Mixtures of Local Experts},
  journal = {Neural Computation},
  volume = {3},
  number = {1},
  pages = {79--87},
  year = {1991}
}

@article{switch,
  author = {Fedus, William and Zoph, Barret and Shazeer, Noam},
  title = {Switch Transformers: Scaling to Trillion Parameter Models with Simple and Efficient Sparsity},
  journal = {Journal of Machine Learning Research},
  volume = {23},
  pages = {1--39},
  year = {2022}
}

@inproceedings{drain,
  author = {He, Pinjia and Zhu, Jieming and Zheng, Zibin and Lyu, Michael R.},
  title = {Drain: An Online Log Parsing Approach with Fixed Depth Tree},
  booktitle = {Proceedings of IEEE International Conference on Web Services},
  year = {2017},
  pages = {33--40}
}

@article{li2022swisslog,
  title={SwissLog: Robust anomaly detection and localization for interleaved unstructured logs},
  author={Li, Xiaoyun and Chen, Pengfei and Jing, Linxiao and He, Zilong and Yu, Guangba},
  journal={IEEE Transactions on Dependable and Secure Computing},
  volume={20},
  number={4},
  pages={2762--2780},
  year={2022},
  publisher={IEEE}
}

@article{le2024prelog,
  title={Prelog: A pre-trained model for log analytics},
  author={Le, Van-Hoang and Zhang, Hongyu},
  journal={Proceedings of the ACM on Management of Data},
  volume={2},
  number={3},
  pages={1--28},
  year={2024},
  publisher={ACM New York, NY, USA}
}

@article{markakis2024logs,
  title={From logs to causal inference: diagnosing large systems},
  author={Markakis, Markos and Youngmann, Brit and Gao, Trinity and Zhang, Ziyu and Shahout, Rana and Chen, Peter Baile and Liu, Chunwei and Sabek, Ibrahim and Cafarella, Michael},
  journal={Proceedings of the VLDB Endowment},
  volume={18},
  number={2},
  pages={158--172},
  year={2024},
  publisher={VLDB Endowment}
}

@article{huang2025no,
  title={No More Labelled Examples? An Unsupervised Log Parser with LLMs},
  author={Huang, Junjie and Jiang, Zhihan and Chen, Zhuangbin and Lyu, Michael},
  journal={Proceedings of the ACM on Software Engineering},
  volume={2},
  number={FSE},
  pages={2406--2429},
  year={2025},
  publisher={ACM New York, NY, USA}
}

@inproceedings{wang2024large,
  title={Large language models can provide accurate and interpretable incident triage},
  author={Wang, Zexin and Li, Jianhui and Ma, Minghua and Li, Ze and Kang, Yu and Zhang, Chaoyun and Bansal, Chetan and Chintalapati, Murali and Rajmohan, Saravan and Lin, Qingwei and others},
  booktitle={2024 IEEE 35th International Symposium on Software Reliability Engineering (ISSRE)},
  pages={523--534},
  year={2024},
  organization={IEEE}
}

@inproceedings{cui2025aetherlog,
  title={AetherLog: Log-based Root Cause Analysis by Integrating Large Language Models with Knowledge Graphs},
  author={Cui, Tianyu and Fu, Ruowei and Liu, Changchang and Ji, Yuhe and Gu, Wenwei and Zhang, Shenglin and Sun, Yongqian and Pei, Dan},
  booktitle={2025 IEEE 36th International Symposium on Software Reliability Engineering (ISSRE)},
  pages={49--60},
  year={2025},
  organization={IEEE}
}

@article{sui2023logkg,
  title={Logkg: Log failure diagnosis through knowledge graph},
  author={Sui, Yicheng and Zhang, Yuzhe and Sun, Jianjun and Xu, Ting and Zhang, Shenglin and Li, Zhengdan and Sun, Yongqian and Guo, Fangrui and Shen, Junyu and Zhang, Yuzhi and others},
  journal={IEEE Transactions on Services Computing},
  volume={16},
  number={5},
  pages={3493--3507},
  year={2023},
  publisher={IEEE}
}

@article{adaptivelog,
  title={Adaptivelog: An adaptive log analysis framework with the collaboration of large and small language model},
  author={Ma, Lipeng and Yang, Weidong and Li, Yixuan and Fei, Ben and Zhou, Mingjie and Li, Shuhao and Jiang, Sihang and Xu, Bo and Xiao, Yanghua},
  journal={ACM Transactions on Software Engineering and Methodology},
  year={2025},
  publisher={ACM New York, NY}
}

@inproceedings{hong2025cslparser,
  title={CSLParser: A Collaborative Framework Using Small and Large Language Models for Log Parsing},
  author={Hong, Weijie and Wu, Yifan and Zhang, Lingzhe and Duan, Chiming and Xiao, Pei and He, Minghua and Yang, Xixuan and Li, Ying},
  booktitle={2025 IEEE 36th International Symposium on Software Reliability Engineering (ISSRE)},
  pages={61--72},
  year={2025},
  organization={IEEE}
}

@inproceedings{le2022log,
  title={Log-based anomaly detection with deep learning: How far are we?},
  author={Le, Van-Hoang and Zhang, Hongyu},
  booktitle={Proceedings of the 44th international conference on software engineering},
  pages={1356--1367},
  year={2022}
}

@inproceedings{rcacopilot,
  author    = {Chen, Yinfang and Xie, Huaibing and Ma, Minghua and Kang, Yu and Gao, Xin and Shi, Liu and Cao, Yunjie and Gao, Xuedong and Fan, Hao and Wen, Ming and Zhu, Jun and Sailer, Avi and Lozano, Levent and Bansal, Chetan and Rajmohan, Saravan and Zhang, Dongmei},
  title     = {Automatic Root Cause Analysis via Large Language Models for Cloud Incidents},
  booktitle = {Proceedings of the Nineteenth European Conference on Computer Systems (EuroSys)},
  year      = {2024},
  pages     = {674--688}
}

@inproceedings{han2024potential,
  title={The potential of one-shot failure root cause analysis: Collaboration of the large language model and small classifier},
  author={Han, Yongqi and Du, Qingfeng and Huang, Ying and Wu, Jiaqi and Tian, Fulong and He, Cheng},
  booktitle={Proceedings of the 39th IEEE/ACM International Conference on Automated Software Engineering},
  pages={931--943},
  year={2024}
}

@INPROCEEDINGS{loghub,
  author={Zhu, Jieming and He, Shilin and He, Pinjia and Liu, Jinyang and Lyu, Michael R.},
  booktitle={2023 IEEE 34th International Symposium on Software Reliability Engineering (ISSRE)}, 
  title={Loghub: A Large Collection of System Log Datasets for AI-driven Log Analytics}, 
  year={2023},
  volume={},
  number={},
  pages={355-366},
  doi={10.1109/ISSRE59848.2023.00071}}

@INPROCEEDINGS{bgl,
  author={Oliner, Adam and Stearley, Jon},
  booktitle={37th Annual IEEE/IFIP International Conference on Dependable Systems and Networks (DSN'07)}, 
  title={What Supercomputers Say: A Study of Five System Logs}, 
  year={2007},
  volume={},
  number={},
  pages={575-584},
  doi={10.1109/DSN.2007.103}}

@inproceedings{bert,
  title={Bert: Pre-training of deep bidirectional transformers for language understanding},
  author={Devlin, Jacob and Chang, Ming-Wei and Lee, Kenton and Toutanova, Kristina},
  booktitle={Proceedings of the 2019 conference of the North American chapter of the association for computational linguistics: human language technologies, volume 1 (long and short papers)},
  pages={4171--4186},
  year={2019}
}

@inproceedings{logbert,
  author    = {Guo, Haixuan and Yuan, Shuhan and Wu, Xintao},
  title     = {{LogBERT}: Log Anomaly Detection via {BERT}},
  booktitle = {Proceedings of the International Joint Conference on Neural Networks},
  year      = {2021}
}

@inproceedings{sun2019bert,
  author    = {Sun, Chi and Qiu, Xipeng and Xu, Yige and Huang, Xuanjing},
  title     = {How to Fine-Tune {BERT} for Text Classification?},
  booktitle = {China National Conference on Chinese Computational Linguistics},
  pages     = {194--206},
  year      = {2019}
}

@misc{openai_pricing,
  author = {{OpenAI}},
  title  = {{Pricing --- {OpenAI} Developer Platform}},
  howpublished = {\url{https://openai.com/api/pricing/}},
  note   = {Accessed: April 2026}
}

@misc{anthropic_pricing,
  author = {{Anthropic}},
  title  = {{Pricing --- {Anthropic} Developer Documentation}},
  howpublished = {\url{https://www.anthropic.com/pricing}},
  note   = {Accessed: April 2026}
}

@misc{google_pricing,
  author = {{Google DeepMind}},
  title  = {{Gemini Developer {API} Pricing}},
  howpublished = {\url{https://ai.google.dev/pricing}},
  note   = {Accessed: April 2026}
}

@INPROCEEDINGS{logparsingeval,
  author={Petrescu, Stefan and Den Hengst, Floris and Uta, Alexandru and Rellermeyer, Jan S.},
  booktitle={2023 IEEE 34th International Symposium on Software Reliability Engineering (ISSRE)}, 
  title={Log Parsing Evaluation in the Era of Modern Software Systems}, 
  year={2023},
  volume={},
  number={},
  pages={379-390},
  doi={10.1109/ISSRE59848.2023.00019}}

@INPROCEEDINGS{logzip,
  author={Liu, Jinyang and Zhu, Jieming and He, Shilin and He, Pinjia and Zheng, Zibin and Lyu, Michael R.},
  booktitle={2019 34th IEEE/ACM International Conference on Automated Software Engineering (ASE)}, 
  title={Logzip: Extracting Hidden Structures via Iterative Clustering for Log Compression}, 
  year={2019},
  volume={},
  number={},
  pages={863-873},
  doi={10.1109/ASE.2019.00085}}

@inproceedings{du2016spell,
  title={Spell: Streaming parsing of system event logs},
  author={Du, Min and Li, Feifei},
  booktitle={2016 IEEE 16th International Conference on Data Mining (ICDM)},
  pages={859--864},
  year={2016},
  organization={IEEE}
}

@article{huang2020hitanomaly,
  title={Hitanomaly: Hierarchical transformers for anomaly detection in system log},
  author={Huang, Shaohan and Liu, Yi and Fung, Carol and He, Rong and Zhao, Yining and Yang, Hailong and Luan, Zhongzhi},
  journal={IEEE transactions on network and service management},
  volume={17},
  number={4},
  pages={2064--2076},
  year={2020},
  publisher={IEEE}
}

@inproceedings{zhang2019robust,
  title={Robust log-based anomaly detection on unstable log data},
  author={Zhang, Xu and Xu, Yong and Lin, Qingwei and Qiao, Bo and Zhang, Hongyu and Dang, Yingnong and Xie, Chunyu and Yang, Xinsheng and Cheng, Qian and Li, Ze and others},
  booktitle={Proceedings of the 2019 27th ACM joint meeting on European software engineering conference and symposium on the foundations of software engineering},
  pages={807--817},
  year={2019}
}

@inproceedings{yang2021plelog,
  title={Plelog: Semi-supervised log-based anomaly detection via probabilistic label estimation},
  author={Yang, Lin and Chen, Junjie and Wang, Zan and Wang, Weijing and Jiang, Jiajun and Dong, Xuyuan and Zhang, Wenbin},
  booktitle={2021 IEEE/ACM 43rd International Conference on Software Engineering: Companion Proceedings (ICSE-Companion)},
  pages={230--231},
  year={2021},
  organization={IEEE}
}

@inproceedings{reimers2019sentence,
  title={Sentence-bert: Sentence embeddings using siamese bert-networks},
  author={Reimers, Nils and Gurevych, Iryna},
  booktitle={Proceedings of the 2019 conference on empirical methods in natural language processing and the 9th international joint conference on natural language processing (EMNLP-IJCNLP)},
  pages={3982--3992},
  year={2019}
}

@inproceedings{makanju2009clustering,
  title={Clustering event logs using iterative partitioning},
  author={Makanju, Adetokunbo AO and Zincir-Heywood, A Nur and Milios, Evangelos E},
  booktitle={Proceedings of the 15th ACM SIGKDD international conference on Knowledge discovery and data mining},
  pages={1255--1264},
  year={2009}
}

@article{xiao2022allinfolog,
  title={AllInfoLog: Robust diverse anomalies detection based on all log features},
  author={Xiao, Ruizhi and Chen, Hao and Lu, Jintian and Li, Weilong and Jin, Shuyuan},
  journal={IEEE Transactions on Network and Service Management},
  volume={20},
  number={3},
  pages={2529--2543},
  year={2022},
  publisher={IEEE}
}

@article{he2021survey,
  title={A survey on automated log analysis for reliability engineering},
  author={He, Shilin and He, Pinjia and Chen, Zhuangbin and Yang, Tianyi and Su, Yuxin and Lyu, Michael R},
  journal={ACM computing surveys (CSUR)},
  volume={54},
  number={6},
  pages={1--37},
  year={2021},
  publisher={ACM New York, NY, USA}
}

@article{dai2020logram,
  title={Logram: Efficient log parsing using $ n $ n-gram dictionaries},
  author={Dai, Hetong and Li, Heng and Chen, Che-Shao and Shang, Weiyi and Chen, Tse-Hsun},
  journal={IEEE transactions on software engineering},
  volume={48},
  number={3},
  pages={879--892},
  year={2020},
  publisher={IEEE}
}

@article{logif,
  title={Unsupervised log message anomaly detection},
  author={Farzad, Amir and Gulliver, T Aaron},
  journal={ICT Express},
  volume={6},
  number={3},
  pages={229--237},
  year={2020},
  publisher={Elsevier}
}

@inproceedings{gururangan2022demix,
  title={Demix layers: Disentangling domains for modular language modeling},
  author={Gururangan, Suchin and Lewis, Mike and Holtzman, Ari and Smith, Noah A and Zettlemoyer, Luke},
  booktitle={Proceedings of the 2022 Conference of the North American Chapter of the Association for Computational Linguistics: Human Language Technologies},
  pages={5557--5576},
  year={2022}
}

@INPROCEEDINGS{11334514,
  author={Qi, Jiaxing and Luan, Zhongzhi and Huang, Shaohan and Fung, Carol and Wang, Yuchen and Wang, Aibin and Zhang, Hongyu and Yang, Hailong and Qian, Depei},
  booktitle={2025 40th IEEE/ACM International Conference on Automated Software Engineering (ASE)}, 
  title={LogMoE: Lightweight Expert Mixture for Cross-System Log Anomaly Detection}, 
  year={2025},
  volume={},
  number={},
  pages={330-341},
  doi={10.1109/ASE63991.2025.00035}
}

@misc{openai_gpt5_2026,
  author = {OpenAI},
  title = {GPT-5 Technical Overview},
  year = {2026},
  howpublished = {\url{https://openai.com}},
  note = {Accessed: April 2026}
}

@misc{anthropic_claude_2026,
  author = {Anthropic},
  title = {Claude Model Documentation},
  year = {2026},
  howpublished = {\url{https://www.anthropic.com}},
  note = {Accessed: April 2026}
}

@misc{google_gemini_2026,
  author = {Google DeepMind},
  title = {Gemini API Documentation},
  year = {2026},
  howpublished = {\url{https://ai.google.dev}},
  note = {Accessed: April 2026}
}
